\documentclass[%
 aip,
 amsmath,amssymb,
 reprint,%
]{revtex4-2}

\usepackage{graphicx}
\usepackage{dcolumn}
\usepackage{bm}
\usepackage{graphicx}

\usepackage[utf8]{inputenc}
\usepackage[T1]{fontenc}
\usepackage{mathptmx}
\usepackage{etoolbox}
\usepackage{hyperref} 
\hypersetup{colorlinks,citecolor=blue, filecolor=blue ,linkcolor=blue , urlcolor=blue, pdftex}

\def \FUW{Institute of Experimental Physics, Faculty of Physics, University of Warsaw, ul. Pasteura 5, 02-093 Warszawa, Poland}
\def \Hefei{Hefei Innovation Research Institute, School of Microelectronics, Beihang University, Hefei 230013, P. R. China}
\def \Strawski{Laboratory of Electrochemistry, Faculty of Chemistry, University of Warsaw, ul. Pasteura 1, 02-093 Warszawa, Poland}
\def \Wroclaw{Department of Semiconductor Materials Engineering, Faculty of Fundamental Problems of Technology, Wrocław University of Science and Technology, Wybrzeże Wyspiańskiego 27, 50-370, Wrocław, Poland}
\def \Joanna{Department of Experimental Physics, Wrocław University of Science and Technology, Wybrzeże Wyspiańskiego 27, 50-37 Wrocław, Poland}
\def \Spain{Geosciences Barcelona (GEO3BCN), CSIC, Lluís Solé i Sabarís
s.n., 08028, Barcelona, Catalonia, Spain}

\makeatletter
\def\@email#1#2{%
 \endgroup
 \patchcmd{\titleblock@produce}
  {\frontmatter@RRAPformat}
  {\frontmatter@RRAPformat{\produce@RRAP{*#1\href{mailto:#2}{#2}}}\frontmatter@RRAPformat}
  {}{}
}%
\makeatother
\begin{document}
\newcommand{\MoS}{$\text{MoS}_{2}$}
\newcommand{\WS}{$\text{WS}_{2}$}
\newcommand{\WSe}{$\text{WSe}_{2}$}
\newcommand{\Xa}{$\text{X}_{1}$}
\newcommand{\Xb}{$\text{X}_{2}$}
\newcommand{\Xc}{$\text{X}_{3}$}
\preprint{AIP/123-QED}

\title[]{Excitonic luminescence of iodine-intercalated HfS$_2$}
\author{N. Zawadzka}
\affiliation{\FUW}
\author{T. Woźniak}
\affiliation{\Wroclaw}
\author{M. Strawski}
\affiliation{\Strawski}
\author{I. Antoniazzi}
\affiliation{\FUW}
\author{M. Grzeszczyk}
\affiliation{\FUW}
\author{K. Olkowska-Pucko}
\affiliation{\FUW}
\author{Z.~Muhammad}
\affiliation{\Hefei}
\author{J. Ibanez}
\affiliation{\Spain}
\author{W. Zhao}
\affiliation{\Hefei}
\author{J. Jadczak}
\affiliation{\Joanna}
\author{R. Stępniewski}
\affiliation{\FUW}
\author{A. Babiński}
\affiliation{\FUW}
\author{M. R. Molas}
\affiliation{\FUW}
\email[Corresponding authors:]{n.zawadzka2@student.uw.edu.pl;adam.babinski@fuw.edu.pl;maciej.molas@fuw.edu.pl}

\date{\today}

\begin{abstract}
Photoluminescence from bulk HfS$_2$ grown by the chemical vapor transport method is reported. 
A series of emission lines is apparent at low temperature in the energy range of 1.4 - 1.5 eV. 
Two groups of the observed excitonic transitions followed by their replicas involving acoustic and optical phonons are distinguished using classical intensity correlation analysis.
The emission is attributed to the recombination of excitons bound to iodine (I$_2$) molecules intercalated between layers of HfS$_2$.
The I$_2$ molecules are introduced to the crystal during the growth as halogen transport agents in the growth process.
Their presence in the crystal is confirmed by secondary ion mass spectroscopy.

\end{abstract}

\maketitle

Although layered transition metal dichalcogenides (TMDs) have been well known and widely investigated for more than 50 years,~\cite{wilson1969} they emerged in the spotlight of material scientists for only over a decade.~\cite{Mak2010} 
The interest is motivated by a variety of properties, which characterize TMDs and the unique dependence of their properties on their thickness in the few-layer limit.
Although about 60 layered TMDs have been recognized to date, researchers have focused mainly their attention on Mo- and W-based semiconducting compounds, which crystallize in hexagonal 2H phase.~\cite{Koperski2017,Wang2018} 
Properties of other TMDs have been less investigated, which opens up new avenues for research.
One of the materials, which recently focused the attention of researchers is hafnium disulfide (HfS$_2$), a member of group IVB TMDs with octahedral coordination (hexagonal 1T, space group: P$\bar{3}$m1), which has recently been shown to exhibit very effective electrical and optoelectronic properties.~\cite{xu2015ultrasensitive, kanazawa2016transistor}
Its calculated carrier mobility at room temperature was shown to  reach up to 1800 cm$^{2}$/Vs.~\cite{Zhang2014}
Moreover, its band alignments have unusually large energies below the vacuum level compared to other TMDs.~\cite{Gong2013, Guo2016} 
This promises an unusual electronic structure of its interfaces with other layered materials which may be important for several applications.~\cite{Wang2018, Zhang2019a, Lukman2020, Zhu2017, Muhammad2022}
The growing interest in HfS$_2$ justifies a need to develop a deeper understanding of its fundamental properties.

To meet the challenge, we report on low-temperature photoluminescence (PL) of the bulk HfS$_2$ grown by the chemical vapor transport (CVT) method.
It is shown that the PL consists of a series of well-resolved lines in the energy range of 1.4 - 1.5~eV. 
The observed emission is attributed to the recombination of bound excitons and their replicas with acoustic and optical phonons.
The excitons are proposed to be bound to neutral iodine (I$_2$) molecules intercalated between layers of the TMD crystal.
The molecules are introduced to the crystal during the growth as halogen transport agents in the CVT growth process.
The presence of iodine in the crystal is confirmed by secondary ion mass spectroscopy (SIMS).
We claim that the presence of intercalated halogen molecules in the CVT-grown HfS$_2$ crystals is more general, as similar PL spectra were also observed in other HfS$_2$ samples including commercially available crystals from Osilla and 2D Semiconductors.


Bulk HfS$_2$ crystals, studied in this work, were synthesized by the CVT method with iodine as a transport agent.~\cite{GrzeszczykPress}
The reaction and growth temperatures were set to 1050~K and 950~K, respectively. The growth was performed continuously for 120~h. With higher stability of HfS$_2$ against iodine, CVT growth starts first at elevated temperatures. 
In the vapour phase, HfI and HfI$_2$ can also dominate the direction of the reaction. 
After 120 h, the reaction was stopped automatically and the reactor was cooled down to room temperature in 5~h. 
High quality HfS$_2$ single crystals of 0.8–1~cm$^2$ size were grown in the low-temperature part of the reactor.
High quality of the grown crystals was confirmed by powder X-Ray diffraction (XRD) measurements found to be in perfect agreement with that of the octahedral 1T phase of HfS$_2$.~\cite{Lucovski1973} 

The PL spectra were measured under laser excitation of $\lambda$=~561~nm (2.21~eV) on samples placed on a cold finger of a continuous flow cryostat. 
The excitation light was focused by means of a 50x long-working distance objective with a 0.55 numerical aperture (NA) producing a spot of about 1~$\mu$m diameter. 
The signal was collected via the same microscope objective (the backscattering geometry), sent through a 0.75~m monochromator, and detected by using a liquid nitrogen cooled charge-coupled device (CCD) camera. 
The excitation power focused on samples was kept at 100~$\mu$W during all measurements to avoid local heating. 

Density functional theory (DFT) calculations were conducted in Vienna Ab initio Simulation Package~\cite{VASP} with Projector Augmented Wave method.~\cite{PAW} 
Perdew–Burke–Ernzerhof parametrization~\cite{PBE} of general gradients approximation to the exchange-correlation functional was used. 
The plane waves basis cutoff energy was set to 550 eV and a 9$\times$9$\times$6 $\Gamma$-centered Monkhorst-Pack k-grid sampling was applied. 
Geometrical structure was optimized with $10^{-5}$ eV/\AA~and 0.01 kbar criteria for the interatomic forces and stress tensor components, respectively. 
Grimme's D3 correction was applied to describe the interlayer vdW interactions.~\cite{D3} 
Spin-orbit interaction was taken into account during geometry optimization. Phonon calculations were performed within Parliński-Li-Kawazoe method,~\cite{Parlinski} as implemented in Phonopy software.~\cite{Phonopy} 
The 3$\times$3$\times$2 supercells were found sufficient to converge the interatomic force constants within the harmonic approximation. 
The phonon density of the states (DOS) and band structure were obtained from integration over full Brillouin zone (BZ) on a 27$\times$27$\times$18 grid.

The in-depth composition of the sample was probed using SIMS. 
As received, samples were transferred without special pre-treatment to the analytical chamber where the pressure was equal to 9x10$^{-10}$ mbar. 
The distribution of elements was obtained with a time of-flight SIMS apparatus (TOF SIMS 5, ION-TOF GmbH, Germany) operating in dual beam mode. 
The samples were sputtered by cesium ions (operating conditions: 1~keV, 76~nA), rastered over 200 $\mu$m x 200 $\mu$m area. 
Exposed this way, internal layers of the sample were analysed with a Bi$^+$ ion beam  (operating conditions: 50 $\mu$m x 50 $\mu$m raster size, 30~keV, 1.12~pA).
The internal mass calibration was performed using mass of ions always present: $^{34}$S$^-$, S$^{2-}$, S$^{3-}$, S$^{4-}$.


\begin{figure}[t]
\includegraphics[width=.48\textwidth]{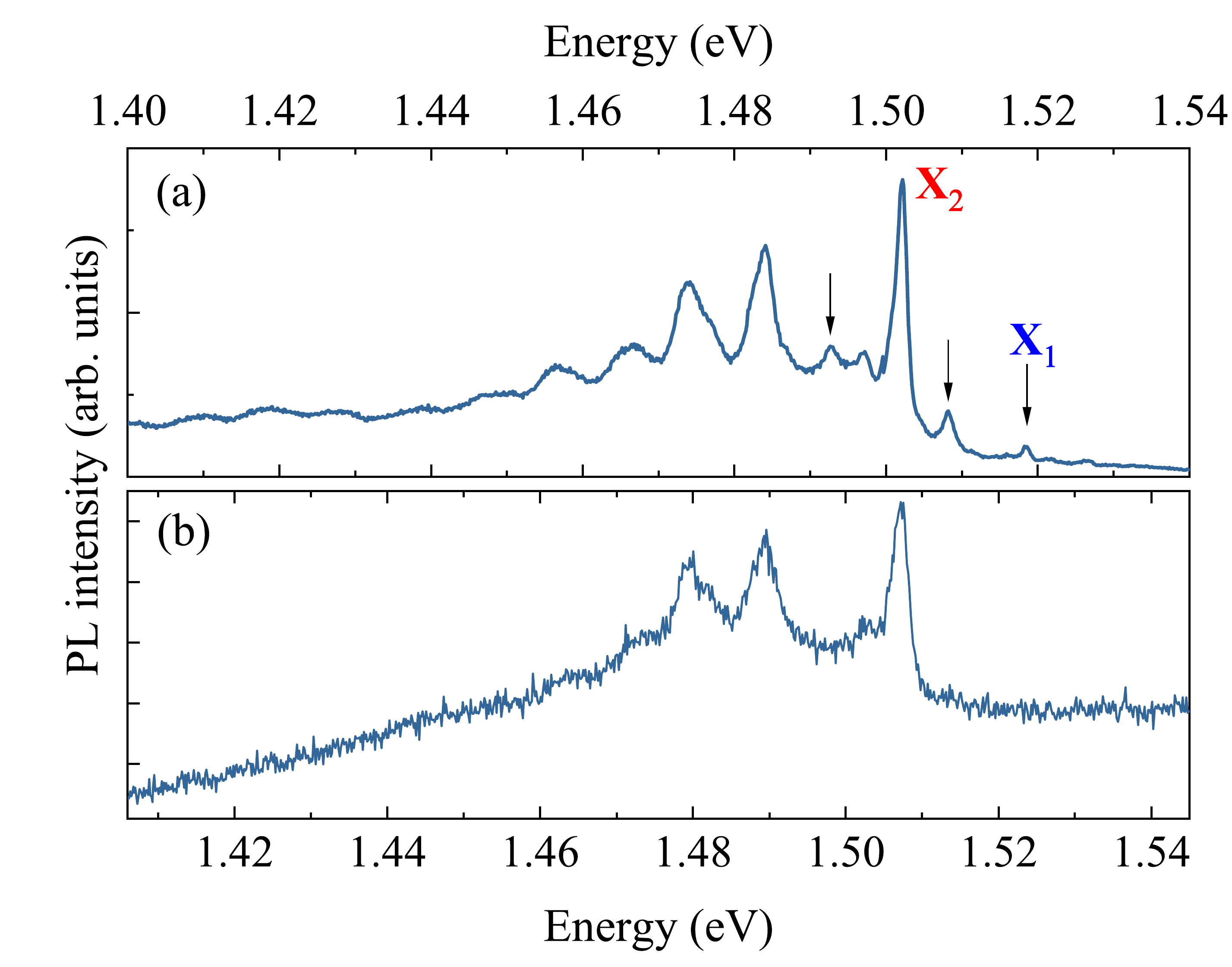}
\caption{
Photoluminescence spectra of HfS$_2$ measured at $T$=5K on a bulk crystal (a) and a flake of 5~nm thickness (b). 
Arrows denote peaks in the spectrum from bulk, which are absent from the flake's spectrum. 
Energy scale of the flake's spectrum is shifted by 5 meV with respect to that of the bulk spectrum.}
\label{bulk_vs_flake}
\end{figure}

The PL spectrum of HfS$_2$ measured at $T$=5~K comprises several emission lines in the energy range 1.4 -- 1.5 eV (see Fig.\ref{bulk_vs_flake}(a)).
The general lineshape of the spectrum does not change over a whole sample.
The detailed analysis of the PL spectra is presented in the Supplementary Material (SM).
Two characteristic peaks: X$_1$ and X$_2$, appear in the spectrum at $E_{\textrm{X}_1}$=1.5184 eV and $E_{\textrm{X}_2}$=1.5021 eV.
Spectrum of a similar lineshape, although blue-shifted by 5 meV was also measured on the HfS$_2$~ flake of approx. 5~nm thickness (see Fig. \ref{bulk_vs_flake}(b)).
Although the spectra are similar, the absence of the X$_1$ peak from the flake's spectrum can be noticed.
Missing are also two other features, denoted in Fig.~\ref{bulk_vs_flake}(a) with arrows.
These lines are red-shifted from X$_1$ by 10.2 meV and 25.7 meV, respectively.
Similar PL spectra were also observed from other CVT-grown HfS$_2$ samples, including commercially available from Osilla and 2D Semiconductors.


\begin{figure}[b]
\includegraphics[width=.385 \textwidth]{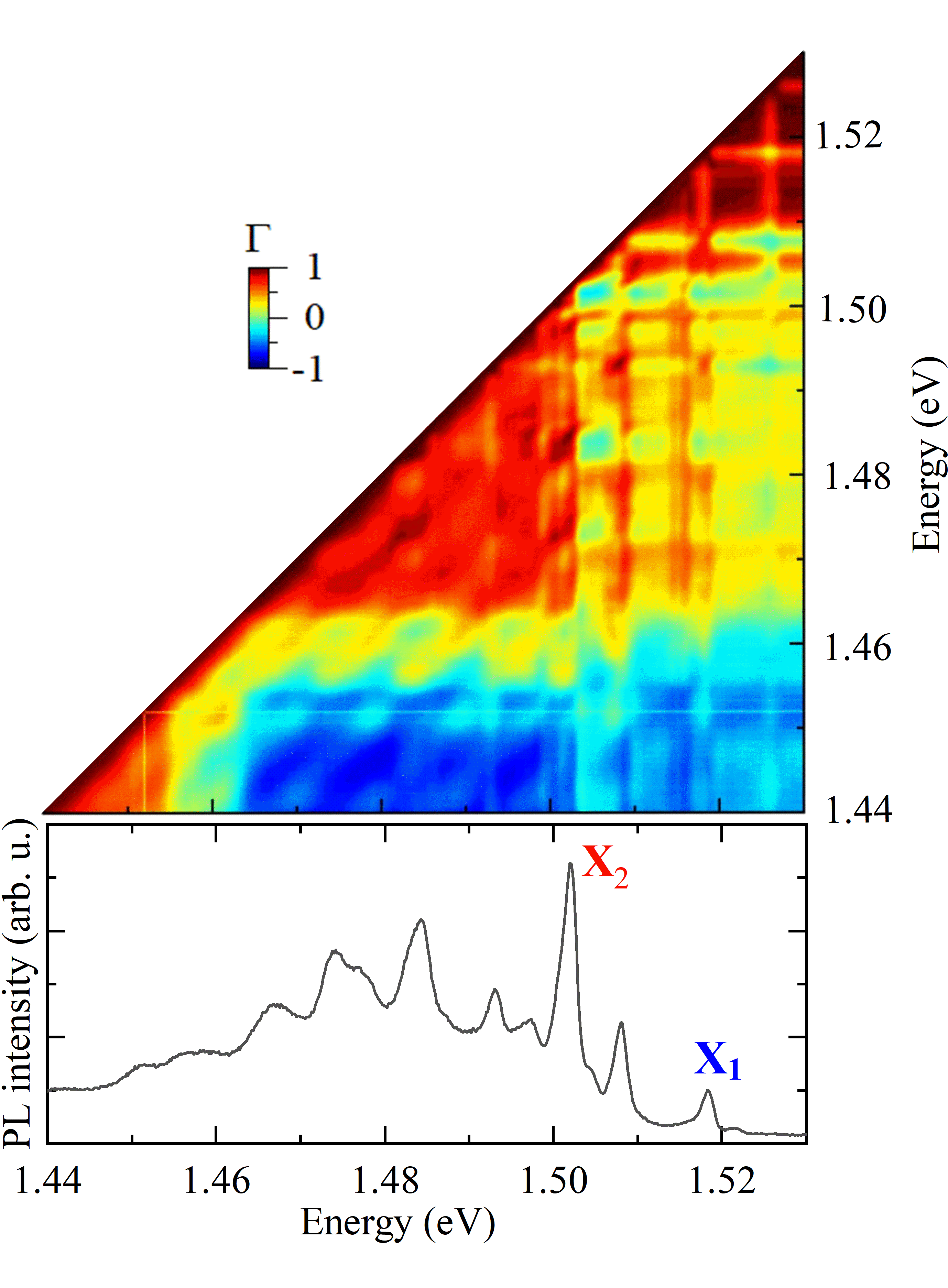}
\caption{
False-color map of the correlation coefficient matrix for the PL spectra from the bulk HfS$_2$~ measured at $T$=5~K. 
A selected PL spectrum is shown in the top panel of the Figure for more clarity. 
The apparent strong correlation signals at LA, TA, LA+TA, and 2TA energy distance from the diagonal can be noticed. }
\label{map}
\end{figure}

The lineshape of the PL spectra suggests their relation to excitonic recombination followed by a series of phonon replicas.
To study the spectra in more detail and profit from their general resemblance over the sample of a millimeter size, 920 spectra were collected from the sample area of approx. 1 mm$^2$ (see SM for more details).
A set of spectra were analyzed and a classical coefficient of the intensity correlation was determined.~\cite{Pietka2013}
The intensity correlation coefficient $\Gamma$ between intensities at different energies $\alpha$ and $\beta$ can be expressed by the formula:
\begin{equation}
    \Gamma=\frac{\Sigma_{i}(I^{\alpha}_{i}-\bar{I^{\alpha}})(I^{\beta}_{i}-\bar{I^{\beta}})}{\sqrt{\Sigma_{i}(I^{\alpha}_{i}-\bar{I^{\alpha}})^2 \Sigma_{i}(I^{\beta}_{i}-\bar{I^{\beta}})^2}},
\end{equation}
where $I^{\alpha}_{i}$ and $I^{\beta}_{i}$ are the PL intensities in the spectrum $i$ measured at energies $\alpha$ and $\beta$, respectively. 
The $\bar{I^{\alpha}}$ and $\bar{I^{\beta}}$ are the intensities averaged over all spectra at energies $\alpha$ and $\beta$, respectively.
The false color map of the $\Gamma$ coefficient for the collected PL spectra (lower panel) together with a typical PL spectrum (upper panel) are shown in Fig.~\ref{map}.
The coefficient exhibits values from -1 to 1, which correspond to deep blue and dark red in the Figure.
The coefficient $\Gamma$=1, apparent on the diagonal of the map, is related to the auto-correlation, $i.e.$ correlation of a given emission line in the spectrum with itself.
Three energy regions can be distinguished in the map with different color patterns.
Most central region is limited by the energies of the X$_2$ emission line at $E_{\textrm{X}_2}$=1.5021 eV and $E$=1.464 eV.
A set of emission lines of high correlation parallel to the diagonal of the map are apparent in the region.
The energy separations between the diagonal and those lines are equal to 10.4 meV, 17.9 meV, 28.1 meV, and 35.3 meV.
For the reason explained later in the text, we refer to them as LA, TA, LA+TA, and 2TA, respectively.
The apparent correlation pattern at energies higher than $E_{X2}$~ is different. 
Lines of high correlation in that energy region are vertical.
Finally, the region of the map with $E$<1.455 eV corresponds to much weaker, low-energy emission, which is most likely related to intrinsic defects in the crystal lattice.~\cite{Kulyuk2003}
This low-energy emission is not addressed in this work.

\begin{figure}[t]
\includegraphics[width=.45\textwidth]{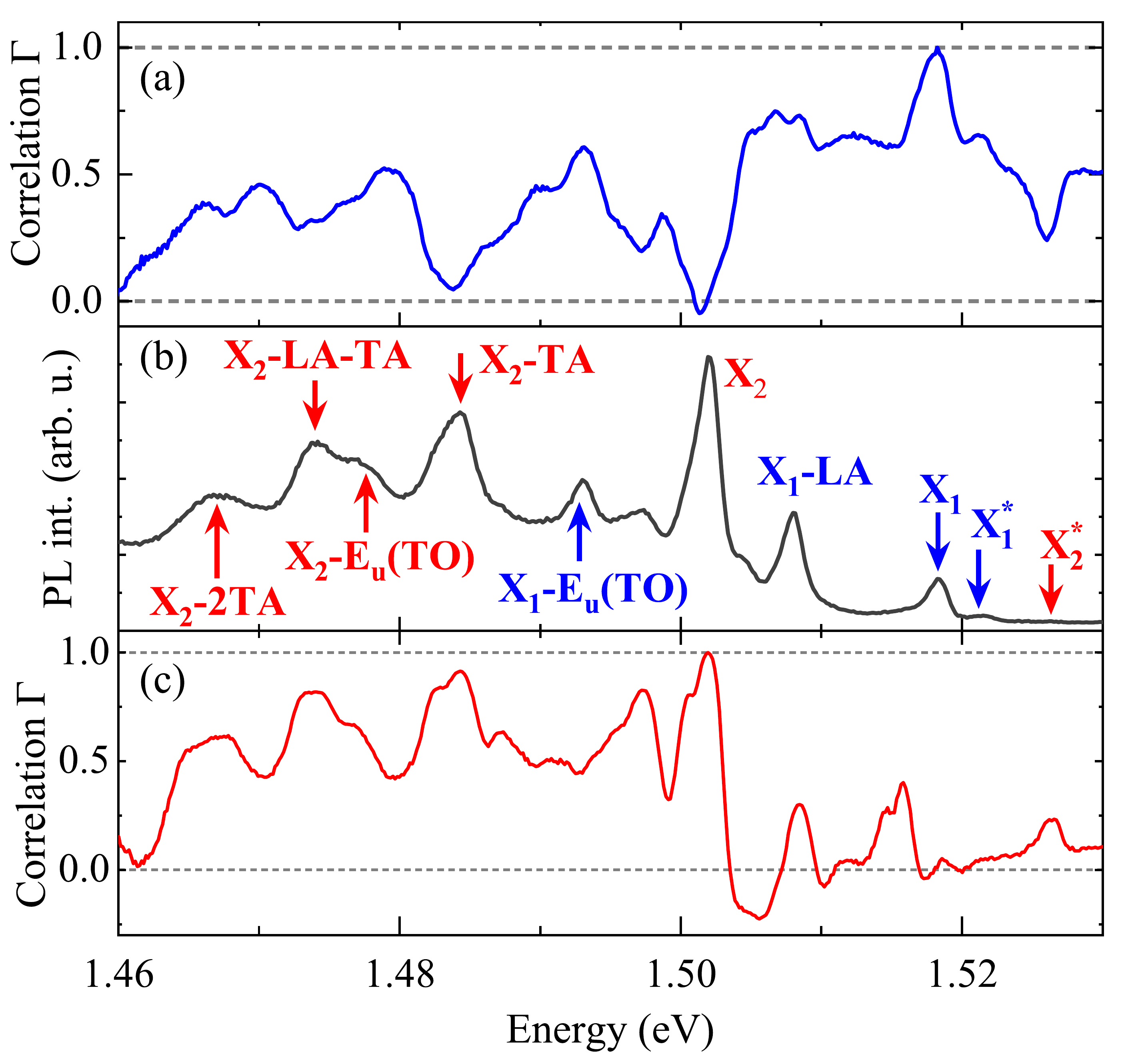}
\caption{
The correlation spectra at the energies of the X$_1$  (a) and X$_2$  (c) emission lines together with a selected PL spectrum (b) plotted for comparison 
The blue arrows indicate emission lines related to the X$_1$  line, while the red arrows correspond to those related to the X$_2$ line. \label{Traces}
}
\end{figure}

Distinct properties of the X$_1$ and X$_2$ emission lines and their phonon replicas can be determined from the correlation matrix.
To see the effect in more detail, we plot the correlation spectra $\Gamma$(E) calculated at the energies of the X$_1$ (panel (a)) and X$_2$ (panel (c)) emission lines, see Fig.~\ref{Traces}.
The PL spectrum (with both X$_1$ and X$_2$ lines) from a selected spot on the sample is also shown in Fig.~\ref{Traces}(b).
As expected, $\Gamma$=1 for the energies of the respective X$_1$ and X$_2$ emission lines, which corresponds to auto-correlation.
Notably, $\Gamma$=1 for X$_1$ (X$_2$) is accompanied by $\Gamma$=0 for X$_2$ (X$_1$).
This confirms the lack of correlation between the presence of X$_1$ and X$_2$ emission lines and can be compared with results shown in Fig.~\ref{bulk_vs_flake}
Both peaks appear in the spectrum independently.
The correlation spectra also reflect a rich structure of replicas that follow the X$_1$ and X$_2$ emission lines.
There are two local maxima in the spectrum for X$_1$ (Fig.~\ref{Traces}(a), which correspond to peaks previously attributed to the phonon replicas of the X$_1$ line.
They are referred to as X$_1$-LA and X$_1$-$E_u$(TO), and their origin will be addressed later.
Similarly, some maxima in the correlation spectrum for X$_2$ (Fig.~\ref{Traces}(c) correspond to peaks apparent in the PL spectra.
This allows to trace the origin of the lines to the X$_2$ line.
The emission lines are referred to as: X$_2$-TA, X$_2$-E$_u$(TO), X$_2$-(LA+TA), and X$_2$-2TA.
There are more local maxima in the correlation spectra (mainly for the X$_1$ line), which cannot be directly correlated with the emission lines in the PL spectra.
The absence of emission features related to those maxima is most likely due to their weak intensity and substantial broadening, which prevents their observation.
One may also appreciate the local maxima of the correlation spectra at energies higher than those of the respective emission line. 
These two lines are referred to as X$_1$$^*$ and X$_2$$^*$ in Fig.\ref{Traces}(b).
As the X$_1$  and X$_2$ lines are associated with the ground-state recombination processes, the X$_1$$^*$ and X$_2$$^*$ emission lines can be related to the corresponding excited states. 

\begin{figure}[b]
	\includegraphics[width=.45\textwidth]{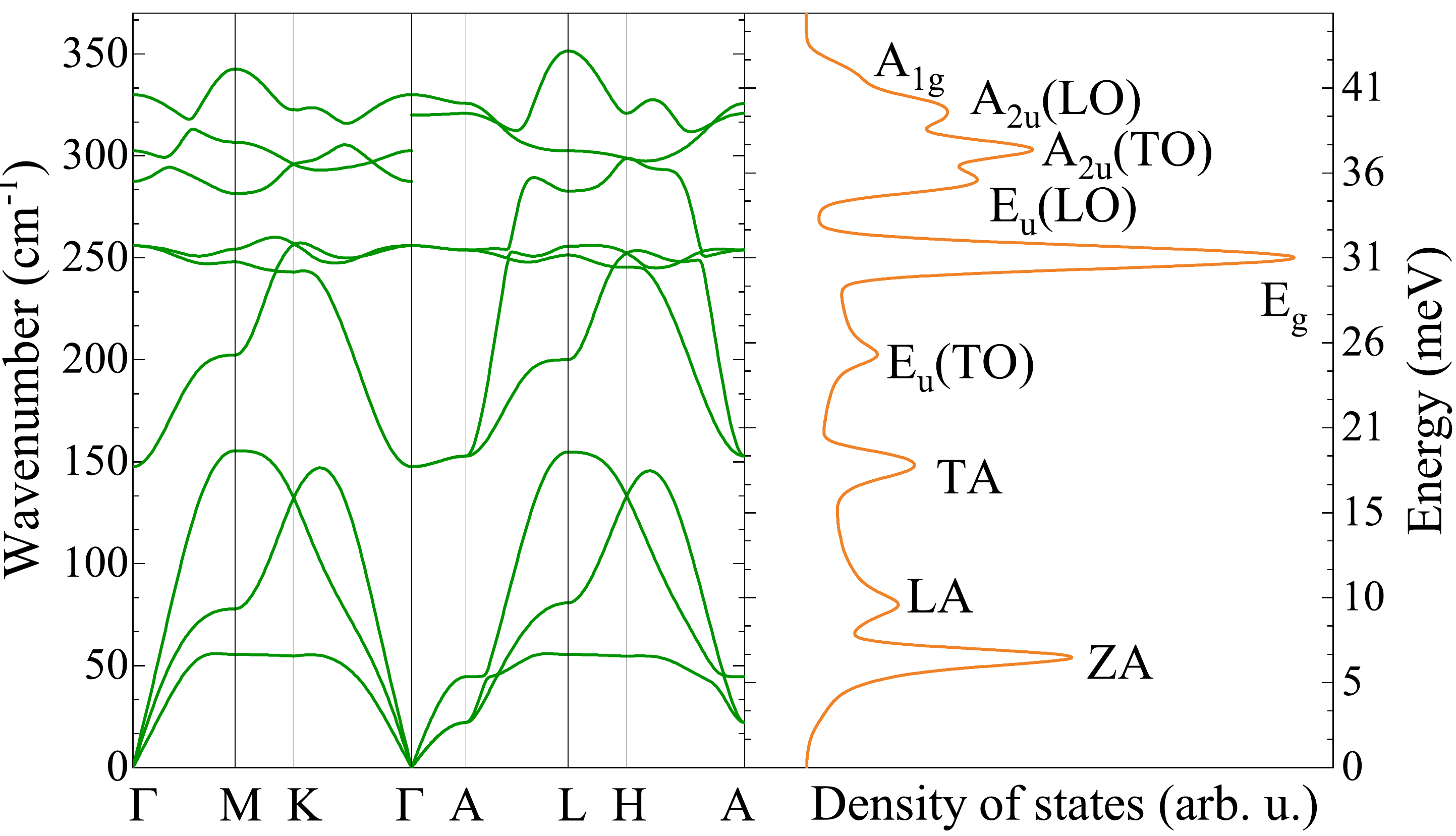}
    \caption{
    The phonon dispersion in bulk HfS$_2$ (a) and the corresponding total integrated phonon density of states obtained by ${ab-initio}$ calculations (b)}
\label{DOS}
\end{figure}

To support the attribution of the observed replicas of the X$_1$ and X$_2$ emission lines to particular phonons in HfS$_2$, DFT calculations were performed.
The resulting phonon dispersion is shown in Fig.\ref{DOS}(a).
There are three branches of acoustic vibrations: longitudinal (LA), transverse (TA), and out-of-plane (ZA) in HfS$_2$, as expected.
Optical modes can be appreciated at energies higher than 150~cm$^{-1}$.
The characteristic LO-TO splitting of the infrared-active E$_u$ and A$_{2u}$ modes can be appreciated.
The former vibrations are active for light's electric field E$\perp$c, while the latter ones are active for E$\|$c, in which c is normal to the layer planes.
The corresponding total phonon DOS is presented in Fig.\ref{DOS}(b).

Three DOS maxima corresponding to acoustic vibrations are referred to as ZA (54~cm$^{-1}$), LA (80~cm$^{-1}$), and TA (148~cm$^{-1}$) after modes that mainly contribute to the DOS at particular energies, see Fig. \ref{DOS}(b).
The DOS maximum at 202 cm$^{-1}$ can be related to the flat regions of the E$_u$(TO) vibration dispersion near M and L points of the BZ in bulk HfS$_2$.
The DOS maximum at 250 cm$^{-1}$ corresponds to the practically dispersion-less Raman-active E$_g$.
The peaks at even higher energy can be related to other infrared- or Raman-active vibrational modes.

\begin{figure}[t]
\includegraphics[width=.45\textwidth]{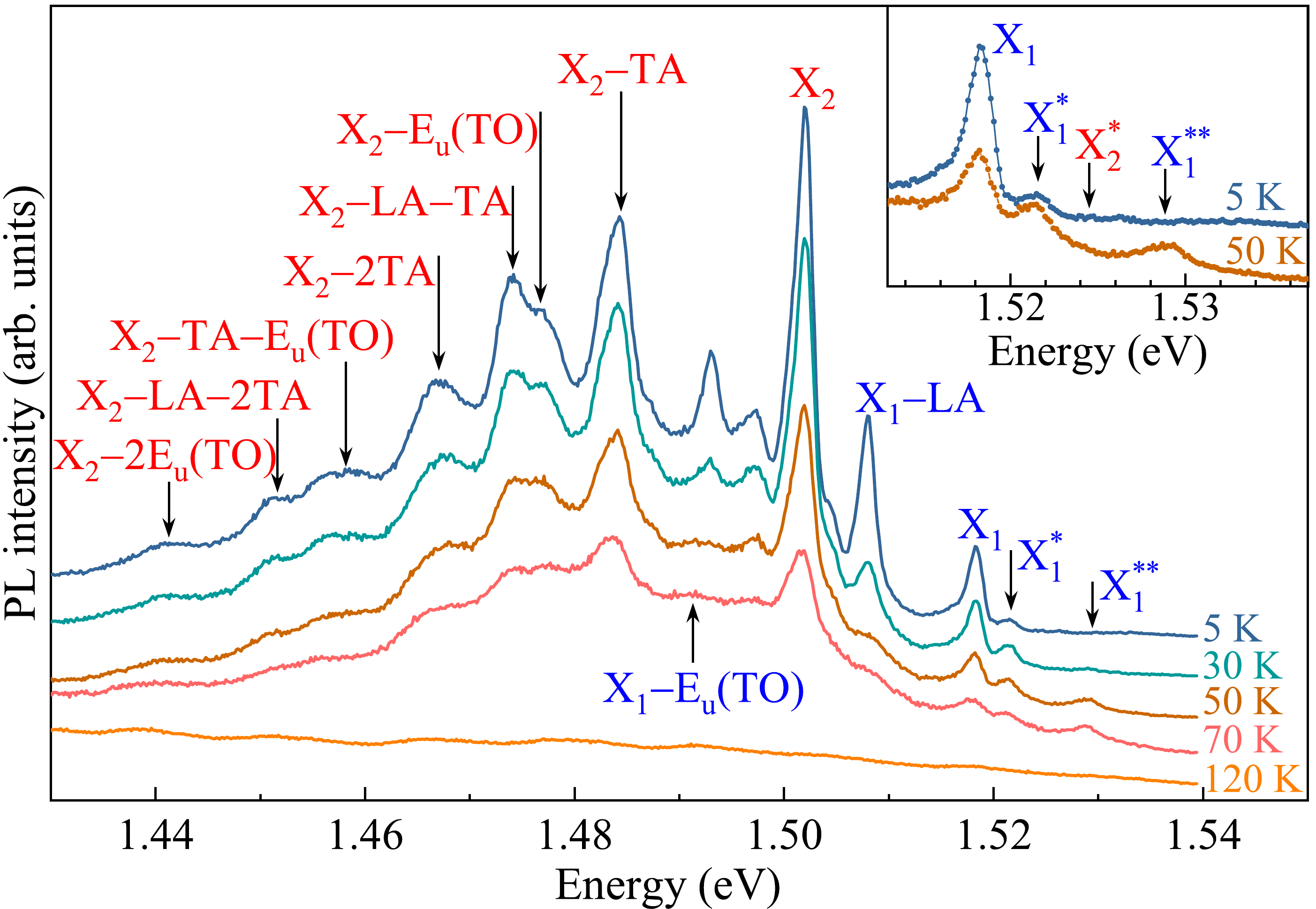}
\caption{
    Photoluminescence spectra at selected temperatures. 
    Energies of several phonon replicas of the excitonic lines are denoted with characteristic phonon modes as determined by DFT calculations.
    The inset displays the spectrum at $T$=5~K and $T$=50~K showing details of the high-energy region of the spectrum. 
  }
\label{fig:1}
\end{figure}

The PL from bulk HfS$_2$ was also examined at higher temperatures - for spectra at selected temperatures see Fig.~\ref{fig:1}.
The emission lines broaden with increasing temperature and eventually quench, leaving the broad-band low-energy emission attributed to intrinsic defects in HfS$_2$.  
The temperature evolution measurements provide yet another proof of the independent origin of the X$_1$ and X$_2$ lines, since the X$_1$ line and its phonon replicas quench at temperatures lower than those of the X$_2$ line and its phonon replicas.
Moreover, two high-energy features, blue-shifted from the X$_1$ line, emerge at elevated temperatures, as can be seen in the inset in Fig.~\ref{fig:1}.
The X$_1^*$~ line (apparent in the correlation spectrum, see Fig.~\ref{Traces}) can be associated with the X$_1$ emission line.
The X$_1^*$~ emission line gains its relative intensity with respect to the X$_1$ line with increasing temperature.
Yet another emission line, denoted X$_1^{**}$, can be distinguished in the Figure, whose attribution to the X$_1$ line is confirmed by correlation analysis at elevated temperature (not shown here).

The expected energy of the X$_2^*$ emission line is also marked in the inset of Fig.~\ref{fig:1}.
It is most likely that at elevated temperatures the X$_2^*$~  line gives rise to the low-energy side-band of the X$_1^{**}$ peak. 

\begin{figure}[t]
	\includegraphics[width=.39\textwidth]{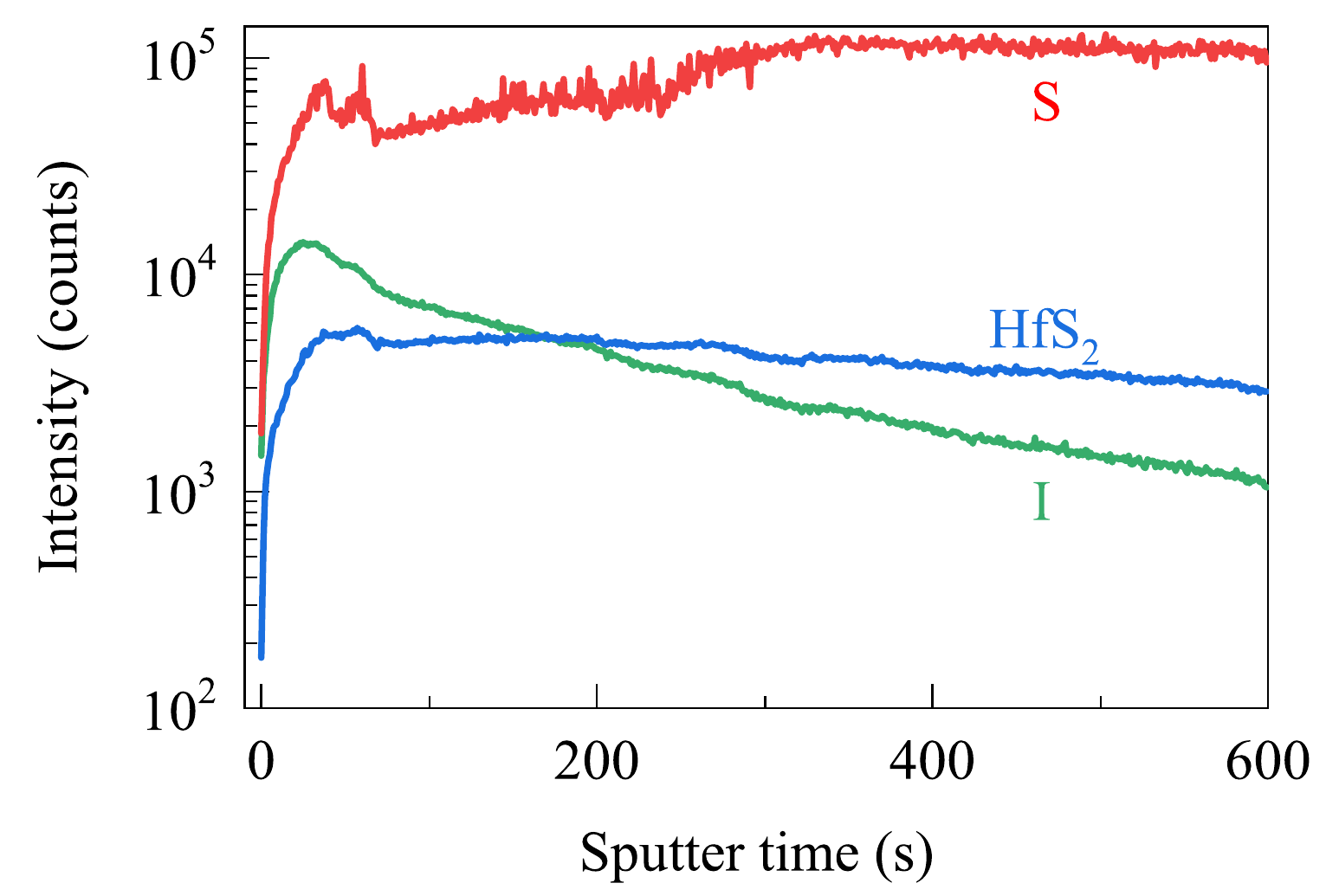}
\caption{The depth profiles of the HfS$_2$, S, and I.}
\label{SIMS}
\end{figure}

The observation of the excitonic PL at approx. 1.5~eV in bulk HfS$_2$, the semiconductor with indirect band gap of approx. 2~eV, might seem surprising.
However, we note that similar low temperature excitonic PL has previously been observed in 2H polytypes of molybdenum/tungsten sulfides/selenides: \MoS:Cl$_2$~\cite{Kulyuk2003}, \WS:Br$_2$~\cite{Kulyuk2005}, \WSe:I$_2$~\cite{Dumchenko2006} or \WS:I$_2$~\cite{Dumchenko2006}.
The reported emission spectra were attributed to excitons bound to intercalated halogen molecules and their phonon replicas with local phonons. 
The ability of TMDs to serve as hosts for intercalating molecules in TMDs originates from their van der Waals (vdW) gaps. 
Each TMD unit cell (layer) comprises planes of covalently bond transition metal atoms "encapsulated" between two chalcogenide atom planes.
The layers are kept together by weaker vdW interactions and separated one from another by the vdW gap. 
The large electron affinity of halogen molecules results in a short range potential attracting electrons from HfS$_2$ layers.~\cite{Kulyuk2003}
The localized electrons interact with optically-excited holes giving rise to bound excitons.
Our results unambiguously point to two distinct excitonic complexes present in the structure: X$_1$ and X$_2$.
Their independent characteristics were confirmed by the accidental absence of the X$_1$ line in the spectrum of the HfS$_2$ flake as well as the intensity correlation analysis (see Figs.~\ref{bulk_vs_flake} and \ref{Traces}).
Moreover, our attribution is supported by the temperature dependence of the spectra with simultaneous quenching of the X$_1$ line and its X$_1$-LA and X$_1$-$E_u$(TO) replicas.
In our opinion, the X$_1$ and X$_2$ emission lines correspond to neutral and charged excitons, respectively.
They combine conduction and valence band carriers in HfS$_2$, which are bound by the iodine molecule. 
Microscopic-scale fluctuations in the unintentional sample doping may be responsible for the observed quenching of the X$_1$ emission line in particular spots of the sample.
Notably, while the intensity of the X$_1$ line fluctuates strongly over the sample area and quenches at some spots, the X$_2$ line is present in all measured spectra. 
This may reflect generally intermediate doping of the investigated sample (allowing for the creation of both neutral and charged complexes) with some more heavily doped spots with trions only.
Assuming the attribution of the observed excitonic lines to the neutral exciton and trion, one can approximate the binding energy of the latter as $\Delta$=16.3 meV.
The attribution of the PL emission to bound excitons comprising band carriers also explains the apparent effect of the sample thinning on their energy.
Thinning the TMD structure results in the energy increase of the conduction band minimum at the Q point of the BZ, which in 2H-TMDs eventually leads to the direct bandgap in the monolayer limit.
A similar shift can be expected in bulk HfS$_2$, which explains the observed evolution of the spectrum with sample thickness.
The attribution of the X$_1$ and X$_2$ lines to bound excitons also facilitates the analysis of their phonon replicas. 
The spatial location of such excitons leads to their delocalization in the momentum space.
Therefore, the exciton can be coupled with phonons of the whole BZ, which explains the rich structure of the phonon replicas in the PL spectrum of the iodine-intercalated HfS$_2$, see Fig.~\ref{bulk_vs_flake}.
If one assumes the coupling of the bound excitons with phonons from the whole BZ, the phonon replica energy structure of the excitonic line should reflect the phonon DOS. 
In particular, the PL features corresponding to the phonon replicas of the X$_1$ and X$_2$ emission lines should be expected at the energies of the phonon DOS maxima.
In fact, the PL peaks related to LA, and E$_u$(TO) replica of the X$_1$ line as well as the TA, E$_u$(TO), LA+TA, and 2TA replica of the X$_2$ line (denoted in Fig.\ref{Traces} with red arrows) can be clearly identified in the PL spectra.
One can appreciate very good agreement of the expected energies with the actual energies of the low-energy satellites of the excitonic features, which supports their attribution to phonon replicas of the excitonic lines.
In fact, most of the observed peaks can be related to the DOS maxima of HfS$_2$~ phonons and not to local vibrations as observed in Refs.~\citenum{Kulyuk2003,Kulyuk2005,Dumchenko2006}.
To confirm our attribution of the observed PL to the halogen species uses as transport agents in CVT growth, SIMS was performed on the investigated crystal.
Figure~\ref{SIMS} presents the evolution of the HfS$_2$, sulfur (S), and iodine (I) concentrations as a function of the sputter time, which can be interpreted as a depth profile of sample. 
The results confirm the presence of iodine in the crystal.
Furthermore, it can be seen that while the HfS$_2$ and S are evenly distributed throughout the crystal, iodine exhibits a higher concentration close to the sample surface and its density is lower in the deeper parts of the sample.
Due to the measured Raman scattering spectrum on the studied HfS$_2$ (see SM for details), we conclude that iodine atoms cannot substitutionally incorporated into the HfS$_2$ layers.
Consequently, as in previously investigated TMD materials,~\cite{Dumchenko2006} it is most likely that the iodine is present in our samples in the form of $I_2$ molecules, which reside in the vdW gaps between HfS$_2$ layers.

In conclusion, the optical emission from the bulk HfS$_2$ is reported. 
A series of well-resolved emission lines, observed at low temperature in the energy range of 1.4 -- 1.5 eV, has been ascribed to bound excitons in HfS$_2$.
Two independent series of excitonic lines followed by acoustic and optical phonon replicas have been identified using a classical analysis of the PL intensity correlations.
It has been proposed that the excitonic lines are due to neutral and charged bound excitons in HfS$_2$.
The excitons are bound by the electron-attractive potential introduced by the I$_2$ molecules intercalated between layers of the crystal.
The I$_2$ molecules are introduced to the crystal during the growth as halogen transport agents in CVT process and their presence in the crystal is confirmed by SIMS.
It is believed that further investigation of the emission will provide important insight in properties of that material and our report would trigger more theoretical studies on possible configurations of $I_2$ molecules in the vdW gaps of HfS$_2$.
More experimental efforts are also necessary to explain the structure of the excited states of the excitonic complexes.   
Moreover, it may be of fundamental importance as similar PL spectra were also observed from other CVT-grown HfS$_2$ samples, including those commercially available.

\section*{Supplementary Material}

See the Supplementary Material for the results of the spatial mapping of the PL spectra made on HfS$_2$ crystal and the analysis of the low-temperature Raman scattering spectrum of the HfS$_2$.

\begin{acknowledgments}
The work has been supported by the National Science Centre, Poland (grant no. 2017/27/B/ST3/00205 and 2018/31/B/ST3/02111).
Z.M. and W.Z. acknowledge support from the National Natural Science Foundation of China (grant no. 62150410438), the International Collaboration Project (no. B16001), and the Beihang Hefei Innovation Research Institute (project no. BHKX-19-02).
DFT calculations were performed with the support of the PLGrid infrastructure.
\end{acknowledgments}

\section*{Data Availability Statement}
The data that support the findings of this study are available from the corresponding author upon reasonable request.

\section*{}
The following article has been accepted by Applied Physics Letters. 
After it is published, it will be found at \href{https://dx.doi.org/10.1063/5.0126894}{doi:10.1063/5.0126894}

\bibliographystyle{apsrev4-2}
\bibliography{biblio}

\newpage
\onecolumngrid
\setcounter{figure}{0}
\setcounter{section}{0}
\renewcommand{\thefigure}{S\arabic{figure}}
\renewcommand{\thesection}{S\Roman{section}}
	\begin{center}
	{\large{{\bf  \textsc{Supplementary Material}} \\ Excitonic luminescence of iodine-intercalated HfS$_2$}}
\end{center}

\section{Spatial mapping of the PL spectra measured on HfS$_2$ crystal}

\begin{figure}[h]
\includegraphics[width=0.95\textwidth]{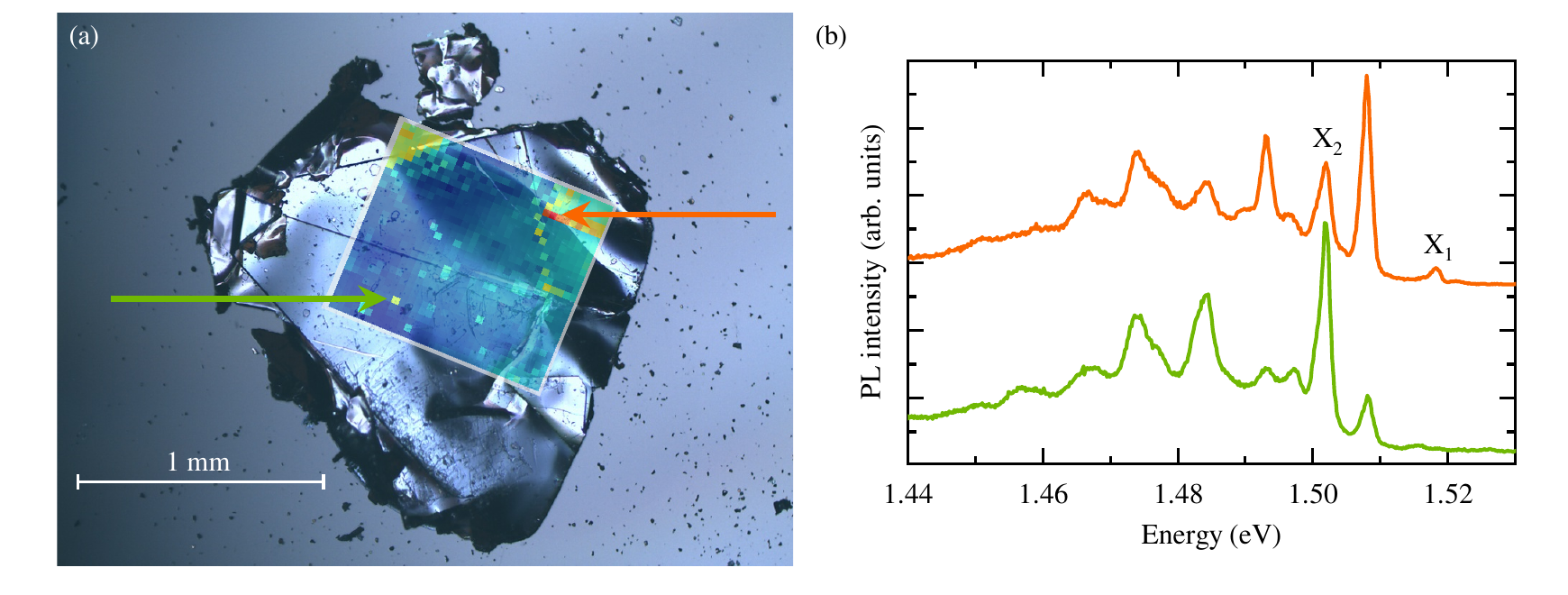}
\caption{(a) The optical microscope image of the HfS$_2$ sample.
    The false-color map, drawn on top of the image, shows the X$_1$/X$_2$ intensity ratio measured at $T$=5~K.
    The size and placement of the map correspond to area under investigation on crystal.
    (b) Two selected low-temperature ($T$=5~K) PL spectra measured on HfS$_2$ at two selected places, marked by colored arrows in panel (a).}
\label{mapka}
\end{figure}

The correlation coefficient map presented in Fig.~2 in the main article was calculated from the PL spectra measured at 920 points in the investigated HfS$_2$ sample.
Fig.~\ref{mapka}(a) presents the optical microscope image of the HfS$_2$ sample.
As is seen from the Figure, the size of the studied HfS$_2$ crystal is of the order of millimeter lateral size. 
The false-color map of the intensity ratio of the zero-phonon lines (X$_1$/X$_2$) is placed on top of the image.
Note that the map corresponds to the spatial area in which 920 PL spectra were measured.
It is seen that the relative intensities stay almost at the same level throughout the investigated area. 
The most spectacular changes in relative intensity can be observed near the places where the surface of the sample is significantly changed, $e.g.$ scratches are seen in the optical image. 
Fig.~\ref{mapka}(b) demonstrates the two selected PL spectra measured at different places on the crystal marked by colored arrows. 
The spectra were chosen to reveal the biggest changes in the X$_1$/X$_2$ relative intensity.
Although the X$_1$ line is well pronounced in the top PL spectrum (orange curve), this line is not visible in the second PL spectrum (green curve).
It confirm that the X$_1$/X$_2$ relative intensity is significantly modified at different places in the HfS$_2$ sample.

\begin{figure}[!h]
\includegraphics[width=.45\textwidth]{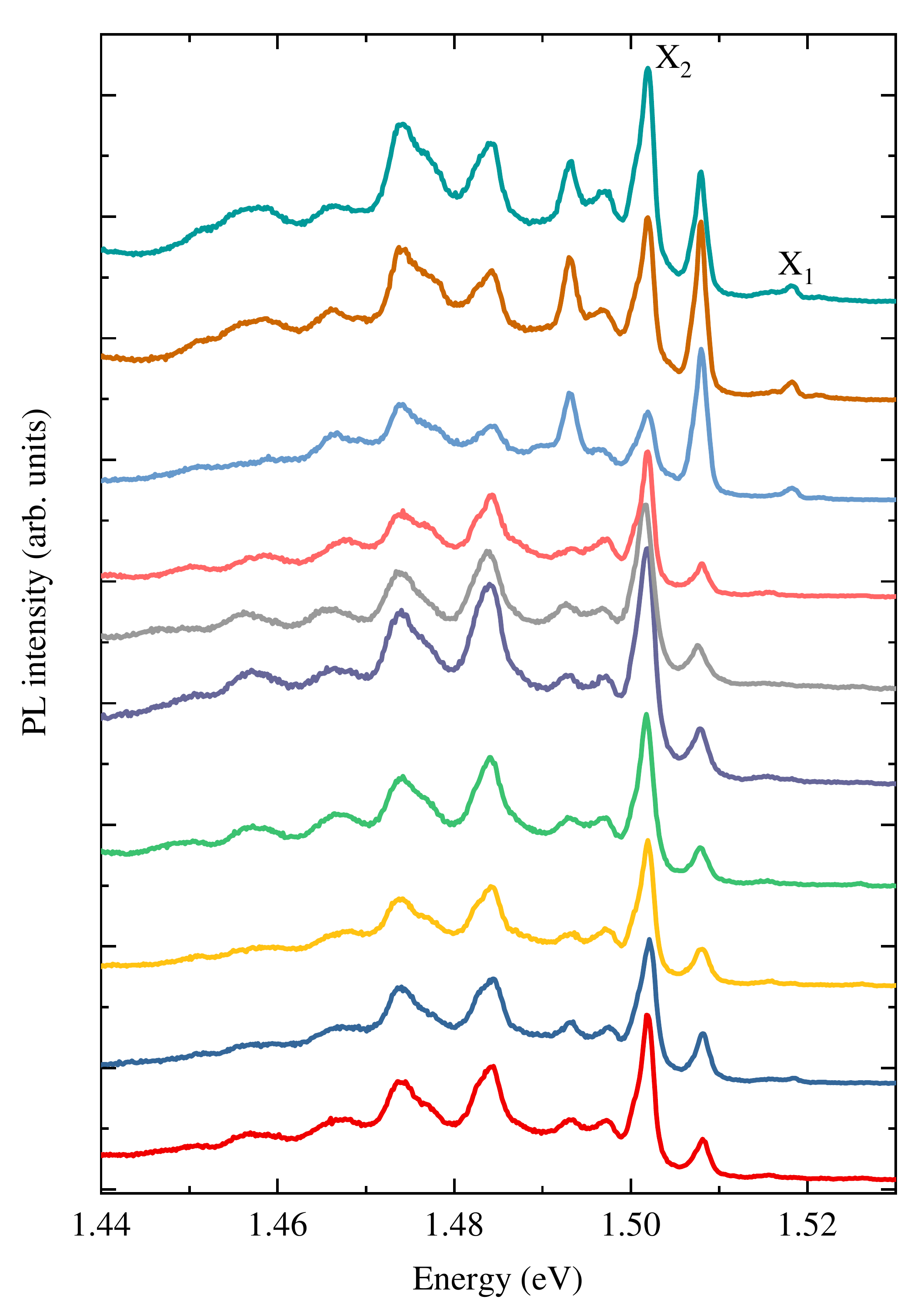}
\caption{Selected low-temperature PL spectra chosen from the measured 920 spectra. 
        Spectra are vertically shifted for clarity.}
\label{PL}
\end{figure}

In order to show that the variation of the X$_1$/X$_2$ relative intensity is accompanied with the corresponding changes in the intensity of the phonon replica lines, we show the low-temperature PL spectra chosen from the measured 920 spectra in Fig.~\ref{PL}. 
It can be concluded that the shape of the low-temperature PL spectra is almost identical for similar X$_1$/X$_2$ relative intensities, while its modifies significantly with the variation of the X$_1$/X$_2$ relative intensity.

\newpage

\section{Low-temperature Raman scattering spectrum measured on HfS$_2$}

\begin{figure}[h]
\includegraphics[width=.45\textwidth]{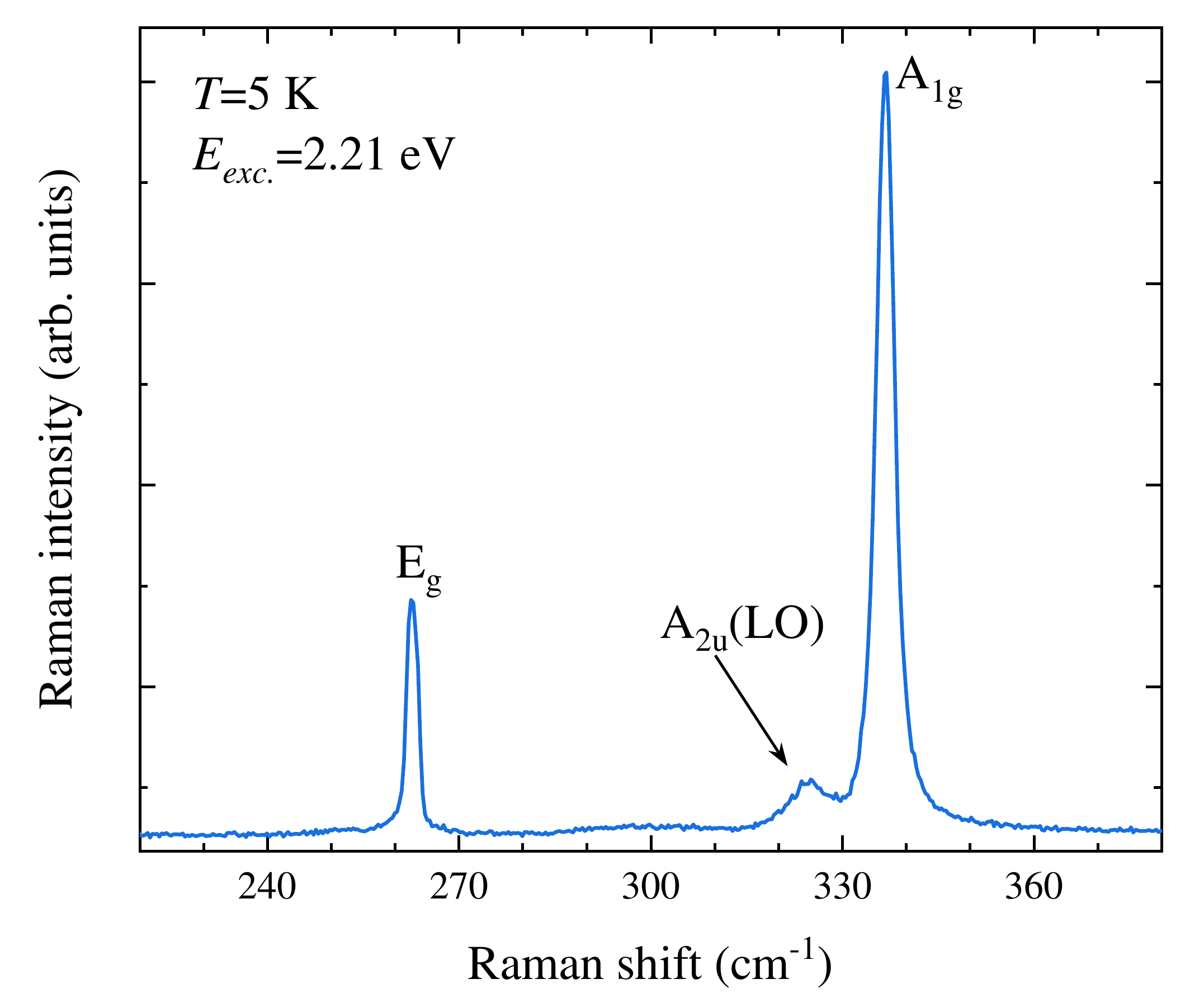}
\caption{Raman scattering spectrum measured on bulk HfS$_{2}$ at $T$=5~K under 2.21~eV laser excitation.}
\label{raman}
\end{figure}

The low-temperature ($T$=5~K) Raman scattering (RS) spectrum of HfS$_{2}$ is presented in Fig.~\ref{raman}.
There are three apparent RS peaks in the spectrum.
According to the literature,~\cite{iwasaki1982Raman, cingolani1987raman, roubi1988resonance, ibanez2018high, neal2021Raman, peng2019, peng2021, Grzeszczyk2022} the peaks that appear at about 260~cm$^{-1}$ and 340 cm$^{-1}$ can easily be attributed to the in-plane E$_\textrm{g}$ and out-of-plane A$_\textrm{1g}$ modes, respectively.
In addition, \mbox{a peak}, distinguished on the low-energy side of the A$_{\textrm{1g}}$ mode, can be ascribed to the A$_{\textrm{{2u}}}\textrm{(LO)}$.\cite{roubi1988resonance}
The full width at half maximum (FWHM) measured for the E$_\textrm{g}$ and A$_\textrm{1g}$ peaks are equal 1.83~cm$^{-1}$ and 3.17~cm$^{-1}$, respectively. 
The obtained FWHMs confirm the high quality of the studied HfS$_2$ crystal.
As a result, the possibility of substitution of Hf and/or S atoms in the layers of HfS$_2$ by iodine atoms is excluded, as substitution of atoms should introduce a significant inhomogeneous broadening of Raman peaks, which is not observed in the RS spectrum.

\end{document}